\documentclass[twocolumn,showpacs,preprintnumbers,amsmath,aps,amssymb]{revtex4-1}
\usepackage{amsmath}
\usepackage[utf8x]{inputenc}
\usepackage{bm}
\usepackage{bigints}
\usepackage{relsize}

   \def\te{{}^3e} 

 \def\dv{\dot v} \def\du{\dot u} \def\dy{\dot y} \def\ham{\mathcal H}
\def\lag{\mathcal L}   
\def\pd{\partial}  
  \def\e{\epsilon}  \def\Om{\Omega} 
 \def\P{\varPi}   \def\b{\beta} 
    
\newcommand{\dep}[2]{\frac{\pd #1}{\pd #2}}

\def\dx{\dot x} \def\dt{\dot t} 
\def\be{\begin{equation*}} \def\te{\end{equation*}} 
\def\bi{\begin{itemize}} \def\ti{\end{itemize}}
\def\ben{\begin{equation}} \def\ten{\end{equation}}
 
\newcommand{\ef}[2]{^{^{#1}\!/\!_{#2}}}

\begin{document}

\title{Conformal Anisotropic Quantum Cosmology}
\author{J. F. Chagoya}
\email{jchagoya@fisica.ugto.mx}
\author{M. Sabido}
\email{msabido@fisica.ugto.mx}
\affiliation{ Departamento  de F\'{\i}sica de la Universidad de Guanajuato,\\
 A.P. E-143, C.P. 37150, Le\'on, Guanajuato, M\'exico
 }%
\date{\today}
\begin{abstract}
In this paper we apply the ideas put forward by Ho\v rava, and introduce anisotropic transformations to cosmology. We start with the Kantowski-Sachs cosmological model and impose anisotropic transformation invariance on the minisuperspace variables. We study the symmetries  of the anisotropic model and by canonical quantization find a Schr\"odinger type equation for  $z\ne 1$. Finally, we conclude that introducing anistropic invariance  can be considered a solution to the problem  of time in quantum cosmology and gives some insight on the structure of a well behaved quantum theory of gravity.
 \end{abstract}
 \maketitle
 \newpage
\section{Introduction}
The search for a Quantum Theory of Gravity has been a long and difficult one, a direct
quantization of General Relativity using the tools of quantum field theory gives a theory
with an ill ultraviolet (UV) behavior. Making matters worst, is the lack of fundamental physical principles to construct the theory.  In \cite{horava1}, by introducing anisotropic scaling transformations
\begin{equation}
t\to b^z t \ \ \  \vec{x}\to b \vec{x}\label{ansc}
\end{equation}
parameterized by a critical exponent $z$, a UV completion to general relativity was proposed. This proposal is known as Ho\v rava's gravity. The attractive feature, was the possible renormalizability  of the gravitational theory at UV fixed point. The price paid for a renormalizable theory of gravity is the loss of Lorentz invariance in the UV limit. The loss of Lorentz invariance is a consequence of the asymmetry in the anisotropic transformations \eqref{ansc} that are enforced. The critical exponent $z$ is taken to have a renormalizable theory in the UV region, but in the infrared (IR) flows to the critical value $z=1$ and Lorentz invariance is recovered. Although this theory has been considered as real candidate for the UV region of GR it is plagued by the existence of (strongly coupled) new degrees of freedom, and for this reason GR cannot be cleanly recovered in the IR \cite{Sotiriou:2010wn}. The cosmological implications have been studied in Ho\v rava gravity by solving the equations of motion for the full theory (see \cite{refs}).

One of the initial approaches to understand quantum gravity, was to work with models with less degrees of freedom. This original approach is named quantum cosmology. The procedure is as follows, one starts with a particular cosmological model an write the Hamiltonian. The quantum version of the theory is obtained by canonically quantizing the theory and is achieved by the usual Dirac quantization program. 
This formalism has the advantage that the inclusion of matter is straight-forward. By considering these models one freezes out degrees of freedom and the canonical quantization of these minisuperspace models gives the Wheeler-DeWitt equation (WDW). The natural question that arises is, can the predictions from quantum cosmology be trusted?. A general analysis  suggest that  conditions can be found to justify the minisuperspace approach and presume the behavior of the wave function as fundamental \cite{halliwell}. Then if quantum cosmology is constructed by paying attention to key features of a full quantum gravity theory, it is likely to capture qualitative features of the full theory.

In this paper we will follow the ideas presented in the previous paragraphs, anisotropic invariance and canonical quantum cosmology, therefore we impose the anisotropic transformations \eqref{ansc}  in the minisuperspace variables. The resulting theory will be an effective theory, we will impose that this theory reduces to the regular GR predictions when the anisotropy is eliminated.
As an example of our proposal we  consider the  Kantowski-Sachs (KS) cosmological model. We will construct a reparametrization invariant cosmological  model, where the minisuperspace variables are compatible with the anisotropic transformations. We study the symmetries of the anisotropic model and find a Schr\"odinger type equation for  $z\ne 1$. In those cases, a first order time derivative appears and a conserved probability current is constructed. In the case $z=1$, the system becomes singular and we return to the original WDW equation. Finally we analyze the cases $z=\pm \infty$ and a linear differential equation is obtained.

The paper is organized as follows, in section II the anistropic  reescaling invariant Lagrangian for the KS model is presented as well as the symmetries of the theory. In Section III, the quantization of the model is performed and the Schr\"odinger type  equation is presented. For a large set of $z$'s we can as well construct the conserved probability current. We exemplify this with
a particular choice of z, however it must remain clear that any $z$ within such set is in principle equally valid.  Finally, section IV is devoted for conclusions and outlook.

\section{The Kantow-Sachs Model}
\subsection{Kantowski-Sachs}

The  Kantowski-Sachs model is one of the most studied anisotropic cosmological models.  Part of the allure, is the wide set of analytical solutions. Recently, research in the KS model  has been carried out in the most diverse manner, i.e. brane world cosmology \cite{cosmos}, scalar field cosmology \cite{cosmo2}, loop quantum cosmology \cite{lqc} to name a few.  In addition to its cosmological relevance , the KS geometry is useful in the description of black holes, a possible connection between KS and quantum black holes has been conjectured  \cite{cosmo3}. Furthermore, the original proposal of noncommutative cosmology was the KS cosmology \cite{ncqc} and it was used to study noncommutative black holes \cite{bh1}.
The  KS model in the Misner parametrization has the metric \cite{misner}
\begin{eqnarray}
 ds^2&=&-N(t)^2dt^2+e^{2\sqrt{3}\b(t)}dr^2+e^{-2\sqrt{3}\b(t)}e^{-2\sqrt{3} \Om(t)}\nonumber\\
&\times&(d\theta^2+\sin^2\theta d\phi^2), \label{ks:metric}
\end{eqnarray}
where $N(t)$ is the lapse function  and $\b(t)$ y $\Om(t)$ are parameters of the metric.
Substituying \eqref{ks:metric} in the Einstein-Hilbert action we get
\begin{equation}
S[\beta,\Om]=\int dt\left[\frac{6A\dot\beta^2}{N}-\frac{6A\dot\Omega^2}{N}+2Ne^{\sqrt 3 \beta} \right],
 \label{ks:accion}
\end{equation}
where we have defined $A=e^{-\sqrt 3 \beta -2\sqrt 3 \Om}$  and the  dot represents a time derivative. 

The original proposal to quantum cosmology was based on canonical quantization of minisuperspace models, where the equation that governs the quantum behavior of the model is the WDW equation.

The  WDW equation for the
KS metric, with some particular factor ordering, is%
\begin{equation}
\left[  -\frac{\partial^{2}}{\partial\Omega^{2}}+\frac{\partial^{2}}%
{\partial\gamma^{2}}+48e^{ -2\sqrt{3}\Omega } \right]
\psi(\Omega,\gamma)=0,\label{ks}
\end{equation}
the solution of this equation is given by \cite{misner},
\begin{equation}
\psi^{\pm}_\nu=e^{\pm i\nu\sqrt{3}\gamma}K_{i\nu}\left(4e^{-\sqrt{3}\Omega}\right) ,
\end{equation}
where $\nu$ is the separation constant and $K_{iv}$ are the modified
Bessel functions. This approach has the problem that wave function is not normalizable and a conserved current density can not be defined. This can be traced to the lack of a first order time derivative on the WDW equation \cite{Unruh:1989db}. In \cite{cosmo3} a possible connection with quantum black holes was conjectured.

For the purpose of introducing anisotropic reescaling,  it is convenient to do the following change of variables in the minisuperspace
$v=e^{-\sqrt 3 \beta}, \ \ \ u=e^{-2\sqrt 3\Om}$
 and the action \eqref{ks:accion} takes the form
\begin{equation}
S[u,v]=\int dt\left[\frac{2A}{N}\frac{\dv^2}{v^2}-\frac{A}{2N}\frac{\du^2}{u^2}+\frac{2N}{v}\right].
 \label{ks:acccionuv}
\end{equation}
We could again calculate the WDW equation in these new variables, but the same problems remain.
\subsection{Anisotropic Invariant Kantowski-Sachs Model.}
Starting with the KS action (\ref{ks:acccionuv}), following \cite{romero} we will write an action invariant under anisitropic reescaling 
 $( u, v, t)\rightarrow(bu, bv, b^zt)$ of the minisuperspace variables. This rescaling will be parametrized by the critical exponent $z$. Evidently  $z=1$ corresponds to a theory in which $u,\ v$ and $t$ transform in the same manner. This value of $z$ correspond to the cosmological model consistent with GR, then we should impose that under anisotropic reescaling the resulting action reduces to action (\ref{ks:acccionuv}). Furthermore, the original action is invariant under time reparametrization, then we will also ask the the new action be time reparametrization invariant.
Imposing the previous requirements we propose
	\begin{eqnarray}
	S(u,v,t)&=&\bigintss d\tau\left[\frac{2A}{N^z}\frac{\dv^{2z}}{v^2\dt^{z-1}}-\frac{A}{2N^z}\frac{\du^{2z}}{u^2\dt^{z-1}}\right. \nonumber\\
	&+&\left. 2
	\dt\left(\frac{N}{\dt v^{2-z}g(z)}\right)^{\frac{z}{2z-1}}
	\right],
	\label{ksa:accion}
	\end{eqnarray}
where the arbitrary  function $g(z)$ has been introduced to assure the anisotropic rescaling invariance, this function takes the value $g(z=1)=1$ and under the anisotropic transformation  $g(z)\rightarrow b^zg(z)$. Furthermore, we have reparemetrized time as $\tau=\tau(t)$, so that time derivatives are given by
$\dot{}=\partial_\tau$.
 In order for this action to be invariant under time reprametrization  $\tau=\tau(f)$ and anisotropic invariance $(u,v,t)\rightarrow (bu,bv,b^z t)$, we have the following conditions
\begin{equation}
	N\rightarrow\dot f N, \ \ \qquad N\rightarrow b^{3-z}N,
\end{equation}
the first condition is the usual transformation in GR and the second one is consistent with the usual gauge choice for  KS cosmology \cite{Barbosa:2004kp}.

\subsection{Canonical Formalism}
The dynamical variables for the action we will be using are $(u,v,t)$ and their corresponding canonical momentum are
\begin{align}
 	&\P_v=\dep{\lag}{\dv}=\frac{4Az(\dv^2)^{z-1}\dv}{N^zv^2(\dt)^{z-1}}, 
	\nonumber \\ 
	&\P_u=\dep{\lag}{\du}=-\frac{Az(\du^2)^{z-1}\du}{N^zu^2\dt^{z-1}}, 
	\nonumber \\
	&\P_t=\dep{\lag}{\dt}=\frac{2A(\dv^2)^z(1-z)}{N^zv^2\dt^z}\nonumber \\
	&\qquad-\frac{A(\du^2)^z(1-z)}{2N^zu^2\dt^z}
	+2\frac{1-z}{1-2z}\left(\frac{N}{\dt v^{2-z}g(z)}\right)^{\frac{z}{2z-1}}.
	\end{align}
The Hamiltonian for this model is given by
	\begin{eqnarray}
	\ham_c&=&\frac{2Az(\dv^2)^z}{N^zv^2\dt^{z-1}}-\frac{Az(\du^2)^z}{2N^zu^2\dt^{z-1}}\nonumber \\ 
	&+&2\dt\frac{z}{1-2z}\left(\frac{N}{\dt v^{2-z}g(z)}\right)^{\frac{z}{2z-1}},
	\label{hamuvt}
	\end{eqnarray}
We can easily see that it can be expressed as
	\begin{equation}
	\ham_c=\frac{\dt z}{1-z}\P_t.
	\label{hampt}
	\end{equation}
The previous equation is ill defined for the KS derived from GR, this is easily seen as it corresponds to $z=1$. This is to be expected because for this value of $z$
in \eqref{ksa:accion} $\dt$ is not present and an associated momentum can not be defined.
\subsection{Infinitesimal Transformations}\label{si}
To simplify the equations of motion we introduce the new coordinates $(x,y)$defined by $(v,u)=(x^p,y^p)$ where $p=\frac{2z}{2z-1}$.
With these new coordinates the anisotropic invariant Lagrangian for the KS model is written as
\begin{equation}
	\lag=\frac{2y^pp^{2z}\dx^{2z}}{N^z\dt^{z-1}}-\frac{x^pp^{2z}\dy^{2z}}{2N^z\dt^{z-1}}
	+2\left(\frac{N}{x^{(2-z)p}g(z)\dt}\right)^{\frac{z}{2z-1}}\dt.
	\label{ksa:lagxy}
	\end{equation}
Varying the action we get the equations of motion
	\begin{align}
	\dot{\P_x}&=\frac{-px^{p-1}p^{2z}\dy^{2z}}{2N^z\dt^{z-1}}
	+2\left(\frac{N}{x^{(2-z)p}g(z)\dt}\right)^{\frac{p}{2}}\frac{p^2(z-2)\dt}{2x}, \nonumber \\
	\dot{\P_y}&=\frac{2py^{p-1}p^{2z}\dx^{2z}}{N^z\dt^{z-1}}, \nonumber \\
	\dot{\P_t}&=0 \nonumber.
	\end{align}
We can se that the equations of motion reduce to the usual GR equations for the KS model \cite{Barbosa:2004kp}  for $z=1$, this is
a consequence of the  way the anistropic invariance was introduced to the Lagrangian \eqref{ksa:lagxy}.
We now proceed to study the symetries of the theory.
The Hamiltonian in the variables $(x,y)$ is
	\begin{eqnarray}
	\ham_c&=&\frac{2zp^{2z}y^p}{N^z\dt^{z-1}}\left[\frac{N^z\dt^{z-1}\P_x}{4zy^pp^{2z}}
	\right]^{p} 
	-\frac{p^{2z}zx^p}{2N^z\dt^{z-1}}\left[\frac{-N^z\dt^{z-1}\P_y}{p^{2z}x^pz}
	\right]^{p}\nonumber\\
	&+&2\left[\frac{N}{x^{(2-z)p}g(z)\dt}\right]^{\frac{p}{2}}\frac{z\dt}{1-2z},
	\end{eqnarray}
The transformation characterized by the parameter $\e(\tau)$ that leaves the action invariant \cite{Henneaux.Teitelboim,romero}, are given by
	\begin{align}
	\{x,\epsilon\ham_c \}&=\epsilon\frac{2zp^{2z}y^p}{N^z}
	\left[\frac{N^z}{y^p4zp^{2z}}\right]^{\frac{2z}{2z-1}}\P_x^{\frac{1}{2z-1}} \nonumber \\
	&=\epsilon\frac{z}{2z-1}\dx, \label{deltax}\\
	\{y,\e\ham_c\}&=-\e\frac{p^{2z}zx^p}{2N^z}
	\left[\frac{-N^z}{p^{2z}zx^p}\right]^{\frac{2z}{2z-1}}\P_y^{\frac{1}{2z-1}} \nonumber \\
	&=\e\frac{z}{2z-1}\dy,\label{deltay} \\
	\{t,\e\ham_c\}&=\frac{z}{1-z}\e , \nonumber \\
	\{\P_x,\e\ham_c\}&=-\e\left\{\frac{p^{2z+1}z\dy^{2z}}{2N^z\dx^{1-p}(2z-1)}+\pd_x\mathcal V\right\} , \nonumber \\
	\{\P_y, \e\ham_c\}&=\e\frac{2zp^{2z}\dx^{2z}p}{N^zy^{1-p}(2z-1)} , \nonumber \\
	\{\P_t,\e\ham_c\}&=0,\nonumber \\
	\delta N(t)&=\pd_\tau\e. \nonumber
	\end{align}
In order to arrive to
 \eqref{deltax} and \eqref{deltay} the explicit form of the momenta is used. These symmetries are consistent with diffeomorphisms on the foliations. We note that the transformations are dependent on $z$ in such a way that they make $\delta t$ transformation invalid for $z=1$. Also the case $z=1/2$ is singular, but unlike the $z=1$ case, there is not any physical meaning to this singularity because it is related to the invalidity of the transformation for $(x,y)$ when $z=1/2$. 
\section{Quantum Formulation}\label{quantum}
In order to construct a  Schr\"odinger type equation for the anisotropic invariant KS cosmological model 
we take equation \eqref{ksa:reldis} and apply canonical quantization
$$\P_v=-\imath\hbar\pd_v, \ \ \ \ \P_u=-\imath\hbar\pd_u, \ \ \ \ \P_t=\imath\hbar\pd_t. $$
The resulting equation is applied to  $\psi(u,v,t)$  and we fix $\dot t=1$ using the invariance under time reparametrization, we get
	\begin{eqnarray}
	\frac{z}{1-z}\imath\hbar\pd_t\psi(u,v,t)&=&\left[\mathcal K(-\imath\pd_v)^{\frac{2z}{2z-1}}
	+\mathcal S(-\imath\pd_u)^{\frac{2z}{2z-1}}\right. \nonumber\\
	&+&\left. \frac{z}{1-2z}\mathcal V\right]\psi(u,v,t),
	\label{ksa:reldis1}
	\end{eqnarray}
where we have defined

\begin{eqnarray}
\mathcal K& =&\left[\frac{N^zv^2h^{2z}}{2Az2^{2z}}\right]^{\frac{1}{2z-1}},	\mathcal S=-\frac{1}{2}\left[\frac{N^zu^2h^{2z}}{Az}\right]^{\frac{1}{2z-1}},\nonumber \\
	\mathcal V&=&2\left[\frac{N}{v^{2-z}g(z)}\right]^{\frac{z}{2z-1}}
\end{eqnarray}
the term $\mathcal V$ is the negative of the KS potential for $z=1$. Using that for  $(q,z_o,z_1)\in\mathbb R$, 
$(-\imath)^{q}=z_0-\imath z_1$, we arrive to the Schr\"odinger type equation
	\begin{eqnarray}
	&&\frac{z}{1-z}\imath\hbar\pd_t\psi(u,v,t)=\left[ (z_0-\imath z_1)
	\mathcal K(\pd_v)^{\frac{2z}{2z-1}}\right.\nonumber \\
	&+&\left.(z_0-\imath z_1)\mathcal S(\pd_u)^{\frac{2z}{2z-1}}
	+\frac{z}{1-2z}\mathcal V\right]\psi(u,v,t).
	\label{ksa:reldis}
	\end{eqnarray} 
The presence of first order time derivative for $t$ is one of the important results of introducing the anisotropic scaling on the minisupersapce variables, this equations is very different to the usual Wheeler-DeWitt equation that appears in canonical quantum cosmology derived from GR, further discussion will be given at the last section.

We are now in a position to  find a continuity equation, for this we follow the usual procedure. We multiply  Eq. \eqref{ksa:reldis} by the complex conjugate of $\psi(u,v,t)$, then take the conjugate of the resulting equations and subtract it to the original one, then we arrive to 
\begin{widetext}
	\begin{eqnarray}
	 \frac{z}{1-z}\imath\hbar(\psi^*\pd_t\psi+\psi\pd_t\psi^*)&=&z_0\mathcal 
	 K(\psi^*\pd_v^{\frac{2z}{2z-1}}\psi
	-\psi\pd_v^{\frac{2z}{2z-1}}\psi^*)-\imath z_1\mathcal K(\psi^*\pd_v^
	{\frac{2z}{2z-1}}\psi 
        +\psi\pd_v^{\frac{2z}{2z-1}}\psi^*) \nonumber\\
        &+&z_0\mathcal S(\psi^*\pd_u^
	{\frac{2z}{2z-1}}\psi -\psi\pd_u^{\frac{2z}{2z-1}}\psi^*)
        -\imath z_1\mathcal S(\psi^*\pd_u^{\frac{2z}{2z-1}}\psi
        +\psi\pd_u^{\frac{2z}{2z-1}}\psi^*).\label{ksa:casicont}
	\end{eqnarray}
\end{widetext}
If we identify the left side of the equation with a time derivative of a probability density, then the right side should be a divergence of a current density. Doing this for an arbitrary value of 
$p\equiv\frac{2z}{2z-1}$ is complicated, but several interesting cases can be worked out. In order to avoid fractional derivatives we will focus our attention for $p\in \mathbb Z$.

\subsection{Anisotropic invariant Kantowski-Sachs Quantum Cosmology for $p\in\mathbb Z$}

First for 
odd values of $p$ we get that  $z_0=0$ and for even values of $p$ we find $z_1=0$. 
\begin{itemize}
 \item $p=0$: both sides of  $\eqref{ksa:reldis}$ vanish, although consistent no conclusions can be made, or any information extracted.
 \item $p=1$: we find that eq. $\eqref{ksa:reldis}$ is linear on the momentum, and corresponds to the case of very large $z$ $(z=\pm\infty)$.
 \item $p=2$: this case corresponds to $z=1$ and is the usual  KS quantum cosmology derived from GR. As expected the derived Schr\"odinger type equation is not valid, because the last term diverges.
 \item $p=3,4$: these cases correspond to  $(z=3/4,2/3,\dots)$, for these cases the this approach to  anisotropic invariant cosmology is valid , and the right side of Eq. \eqref{ksa:casicont} can be written as a divergence, therefore a current density can be defined. 	
\end{itemize}

As an example we present the case  $p=4$. The anisotropic invariant Lagrangian for the KS model for $p=4$ is 
	\begin{equation}
	L=-\frac{A}{2N\ef{2}{3}}\frac{\du\ef{4}{3}}{u^2\dt\ef{-1}{3}}+2\left(\frac{N}{\dt v\ef{4}{3}}\right)^2\frac{\dt}{[g(z)]^2}.
	\end{equation}
We calculate the momenta but in the variables $(x,y)\equiv(v^{1/4},u^{1/4})$ and get

	\begin{eqnarray}
 	\P_x&=&\dep{\lag}{\dx}=\frac{4\ef{11}{6}}{N\ef{2}{3}}\frac{4}{3}(\dx\dt)\ef{1}{3}y^4,\nonumber\\ 
	 \P_y&=&\dep{\lag}{\dy}=-\frac{4\ef{4}{3}}{2N\ef{2}{3}}\frac{4}{3}(\dy\dt)\ef{1}{3}x^4, 
	\nonumber\\
	\P_t&=&\dep{\lag}{\dt}=\frac{2}{3}\frac{(4\dx)\ef{4}{3}}{N\ef{2}{3}}y^4\dt\ef{-2}{3}
	-\frac{1}{3}\frac{(4\dy)\ef{4}{3}}{2N\ef{2}{3}}x^4\dt\ef{-2}{3}\nonumber\\
	&-&2\left(\frac{N}{x\ef{16}{3}}
	 \right)^2\frac{1}{\dt^2[g(z)]^2}.
	\end{eqnarray}
Following the presented formalism, is easy to verify that 	$\ham_c=2\dt\P_t.$
Furthermore, introducing in the Hamiltonian the expressions for
 $\dx$ y $\dy$ as function of the momenta and $\dt$ 
we get
	\begin{align}
	2\dt\P_t&=\frac{3^3N^2}{4^7y^{12}\dt}\P_x^4-\frac{6^32N^2}{16^4x^{12}\dt}\P_y^4
	-4\left(\frac{N}{\dt x\ef{16}{3}}\right)^2\frac{\dt}{[g(z)]^2},
	\end{align}
Now we apply the usual representation for the momenta $(\P_x,\P_y,\P_t)$ and take 
$\dt=1$ we arrive to the Schro\"dinger type equation
	\begin{equation}
	\imath\hbar\pd_t\psi=-\kappa\frac{N^2}{y^{12}}\pd_x^4\psi-\sigma\frac{N^2}{x^{12}}\pd_y^4\psi-V,
	\end{equation}
where $\kappa=-\frac{1}{2}\frac{3^3\hbar^4}{4^7}$, $\sigma=\frac{6^3\hbar^4}{16^4}$ and 
$V=\mathcal V$. Following the method already presented we calculate the continuity equation
	\begin{eqnarray}
	\imath\hbar\pd_t(\psi^*\psi)&=&-\kappa\frac{N^2}{y^12}(\psi^*\pd_x^4\psi-\psi\pd_x^4\psi^*)\nonumber \\
	& &-\sigma\frac{N^2}{x^12}(\psi^*\pd_y^4\psi-\psi\pd_y^4\psi^*).
	\end{eqnarray}
By inspecting the continuity equations and  by a direct calculation we can define a density current
	\begin{equation}
	J_x=\frac{\alpha N^2}{\imath\hbar y^12}j_x, \ \ \ \ \ J_y=\frac{\sigma N^2}{\imath\hbar x^12}j_y,
	\end{equation}	
and finally arrive to the continuity equation in terms of a probability density and current, for the KS anisotropic invariant quantum cosmology with  $z=2/3$
	\begin{equation}
	\pd_t(\psi^*\psi)+\pd_xJ_x+\pd_yJ_y=\pd_t(\psi^*\psi)+\nabla\cdot\vec J=0.
	\end{equation}
Another interesting case is $p=1$ (which correspods to
$z\to\pm\infty$). For $z\to-\infty$, starting from \eqref{ksa:reldis1} and the definitions for $\mathcal K$, $\mathcal S$ y $\mathcal V$ we get the following behaviour for the Schr\"odinger type equation:
	\begin{equation}
	-\imath\hbar\pd_t\psi=-\frac{N^{1/2}}{2}\imath\hbar\pd_v\psi+\frac{N^{1/2}}{2}\imath\hbar\pd_u\psi
	-(\text{ $\propto v^{-\infty}$}),
	\label{ksa:wdzminfty}
	\end{equation}
we also analyze  $z\to\infty$ and find:
	\begin{equation}
	-\imath\hbar\pd_t\psi=-\frac{N^{1/2}}{2}\imath\hbar\pd_v\psi+\frac{N^{1/2}}{2}\imath\hbar\pd_u\psi
	-(\text{ $\propto v^{\infty}$}).
	\label{ksa:wdzinfty}
	\end{equation}
We have assumed that  $g(z\to\pm\infty)$is bounded and has no zeros. It is possible have a $g(z)$ that gives a finite value for the las terms in  \eqref{ksa:wdzminfty} and
\eqref{ksa:wdzinfty}, but then $g$ will be a function of $v$, and this will modify the equations of motion. Then in order to bound the last term of  \eqref{ksa:wdzminfty} 
we need that $v>1$ and for $\eqref{ksa:wdzinfty}$ we have that $v<1$.

\section{Conclusion}

In this paper we present a proposal to introduce anisotropic reescaling invariance in quantum cosmology, this approach is different but inspired in Horava gravity. The presentation is given in terms of an example, KS cosmology.
In particular the loss of Lorentz invariance results in momenta $\Pi_t$ in the Hamiltonian, that after canonical quantization gives a linear time derivative. Also instead of a WDW equation we get a Schr\"odinger type equation. This might indicate the the introduction of this anisotropy could give a better quantum behavior to the theory as in Horava gravity.

As a consequence of the manner we introduce the anisotropy, the limit $z=1$ is well defined, that means that we arrive to KS cosmological model derived from GR, unlike with quantum gravity at a Lifshitz Point (QGLP) where it is not clear that for  $z\to 1$ we get GR. In the new model diffeomorphism invariance is maintained  but it is reflected in the time reparametrization invariance of the cosmological model. 
In the quantum model the WDW equation is obtained for $z=1$, this is to be expected  as in this case there is no anisotropic rescaling. However for other values of $z$ a first order time derivative appears in the quantum equation, from which a probability density and current are constructed. It is worthwhile to remark the difference between a classical limit and the
 limit to GR, the former can be obtained from the quantum equations for any $z$ by using a standard WKB formalism and the result agrees with the already constructed classical equations of motion, nevertheless such classical models are unphysical as we know that Lorentz invariance
is well tested in the infrared; on the other hand, the GR limit is obtained only for $z=1$. 

Finally we conclude that introducing anistropic invariance in the minisuperspace action, we have a proposal that in some sense solves the problem of time in quantum cosmology and furthermore it might give some insight on the structure of a well behaved quantum theory of gravity. So far, there are no direct observational guides for the construction of quantum
 cosmological models ( although there might be in a not-so-far future), but in
the current situation, a minimal and common requirement for these models is that the known
 infrared physics might be recovered, in this sense our proposal is well behaved.  Non the less, there are issues that should be resolved, for example the appearance of fractional order derivatives on the resulting quantum equation. These and other issues are under research and will be reported elsewhere.

\section*{Acknowledgments}
J.C is supported by CONACyT graduate grant.  M. S. is partially supported by CONACYT grants 62253, 135023 and DAIP 125/11.


\begin{thebibliography}{cc}

\bibitem{horava1} P.~Horava,
  Phys.\ Rev.\ D {\bf 79} (2009) 084008.
\bibitem{Sotiriou:2010wn} 
  T.~P.~Sotiriou,
  J.\ Phys.\ Conf.\ Ser.\  {\bf 283}, 012034 (2011);O.~Bertolami and C.~A.~D.~Zarro,
  Phys.\ Rev.\ D {\bf 84}, 044042 (2011).
\bibitem{refs} S.~Mukohyama,
  Class.\ Quant.\ Grav.\  {\bf 27}, 223101 (2010);
  S.~K.~Chakrabarti, K.~Dutta and A.~A.~Sen,
  Phys.\ Lett.\ B {\bf 711}, 147 (2012).

\bibitem{halliwell} J.J. Halliwell, Proc. 13th. Int. Conf. on General Relativ- ity, ed R.J. Gleisser, C.N. Kozameh and O.M. Moreschi (Bristol: IOP Publishing, 1993).
\bibitem{cosmos} 
  A.~N.~Makarenko, V.~V.~Obukhov and K.~E.~Osetrin,
  gr-qc/0301124;  L.~A.~Gergely,
  Class.\ Quant.\ Grav.\  {\bf 21}, 935 (2004).
\bibitem{cosmo2} 
  D.~M.~Solomons, P.~Dunsby and G.~Ellis,
  Class.\ Quant.\ Grav.\  {\bf 23}, 6585 (2006).
  \bibitem{lqc} L.~Modesto,
  Int.\ J.\ Theor.\ Phys.\  {\bf 45}, 2235 (2006).
\bibitem{cosmo3} 
  O.~Obregon and M.~P.~Ryan,
  Mod.\ Phys.\ Lett.\ A {\bf 13}, 3251 (1998); A.~Ashtekar and M.~Bojowald,
  Class.\ Quant.\ Grav.\  {\bf 23} (2006) 391.
\bibitem{ncqc} H.~Garcia-Compean, O.~Obregon and C.~Ramirez,
  Phys.\ Rev.\ Lett.\  {\bf 88}, 161301 (2002).
\bibitem{bh1}
  J.~C.~Lopez-Dominguez, O.~Obregon, M.~Sabido and C.~Ramirez,
  Phys.\ Rev.\ D {\bf 74}, 084024 (2006). 
\bibitem{misner} C. Misner, "Minisuperspace", in {\it Magic without Magic: %
  John Archibald Wheeler} (Freeman, 1972).
  \bibitem{Unruh:1989db} 
  W.~G.~Unruh and R.~M.~Wald,
  Phys.\ Rev.\ D {\bf 40}, 2598 (1989).

  \bibitem{romero} J.~M.~Romero, V.~Cuesta, J.~A.~Garcia and J.~D.~Vergara,
  Phys.\ Rev.\ D {\bf 81}, 065013 (2010).
\bibitem{Barbosa:2004kp} 
  G.~D.~Barbosa and N.~Pinto-Neto,
  Phys.\ Rev.\ D {\bf 70}, 103512 (2004)
\bibitem{Henneaux.Teitelboim} M. Henneaux and C. Teitelboim,
  ``Quantization of gauge systems'' (Princeton University Press,
  1992).


\end{thebibliography}
\end{document}